# cuTT: A High-Performance Tensor Transpose Library for CUDA Compatible GPUs

**ANTTI-PEKKA HYNNINEN**, NVIDIA Corporation, Santa Clara, CA 95050
**DMITRY I. LYAKH**, National Center for Computational Sciences, Oak Ridge National Laboratory, Oak Ridge TN 37831

We introduce the CUDA Tensor Transpose (cuTT) library that implements high-performance tensor transposes for NVIDIA GPUs with Kepler and above architectures. cuTT achieves high performance by (a) utilizing two GPU-optimized transpose algorithms that both use a shared memory buffer in order to reduce global memory access scatter, and by (b) computing memory positions of tensor elements using a thread-parallel algorithm. We evaluate the performance of cuTT on a variety of benchmarks with tensor ranks ranging from 2 to 12 and show that cuTT performance is independent of the tensor rank and that it performs no worse than an approach based on code generation. We develop a heuristic scheme for choosing the optimal parameters for tensor transpose algorithms by implementing an analytical GPU performance model that can be used at runtime without need for performance measurements or profiling. Finally, by integrating cuTT into the tensor algebra library TAL-SH, we significantly reduce the tensor transpose overhead in tensor contractions, achieving as low as just one percent overhead for arithmetically intensive tensor contractions.

## 1. INTRODUCTION

Tensors are mathematical objects that are ubiquitous in many scientific disciplines, including electronic structure theory [Bartlett & Purvis 1980], [Bartlett & Musial 2007], [Lyakh et al 2012], [Lyakh & Bartlett 2014], [Chan et al 2016], [Nakatani & Chan 2013], [Lyakh 2014] and nuclear structure theory [Hagen et al 2014], [Signoracci et al 2015], where they represent the many-body wave-function, quantum information science [Wang et al 2015] and quantum gravity theory [Bao et al 2015], where they encode quantum entanglement, deep learning [Goldsborough 2016] and multivariate data analytics [Austin et al 2016], where tensors reflect complex relations between objects, and bioinformatics, where tensors encode hypergraphs [Weighhill & Jacobson 2015], for example. In our discussion, we adopt the most general definition of a tensor as a multidimensional array of data. In practice, numerical tensor algebra is mostly based on a few numeric primitives, among which the tensor contraction operation is perhaps the most important one, being a generalization of the vector-vector, matrix-vector, and matrix-matrix products from the basic linear algebra. Since a tensor contraction is essentially a partial (or full) reduction over one or more pairs of dimensions in the input tensors, it is often characterized by high compute (arithmetic) intensity, thus favoring modern many-core computer architectures. Consequently, there has been a lot of interest in recent years focused on an efficient implementation of the tensor contraction operation on shared- and distributed-memory computer platforms, including multicore [Epifanovsky et al 2013], [Springer & Bientinesi 2016] and accelerated node architectures (e.g., the TAL-SH library presented in this work) as well as large-scale HPC systems [Baumgartner et al 2005], [Solomonik et al 2013], [Rajbhandari et al 2014a,b], [Calvin et al 2015]. Since the dense tensor contraction operation may still be sensitive to the communication/memory bandwidth, the underlying optimizations in particular included communication avoiding and communication pattern

Author's addresses: Antti-Pekka Hynninen (ahynninen@nvidia.com), NVIDIA Corporation, Santa Clara CA 95050, USA, Dmitry I. Lyakh (liakhdi@ornl.gov), National Center for Computational Sciences, Oak Ridge National Laboratory, Oak Ridge TN, 37831, USA.

regularization [Solomonik et al 2013], [Rajbhandari et al 2014a,b], as well as an optimization of the memory access pattern on individual compute devices [Springer et al 2016a,b], [Springer & Bientinesi 2016], [Matthews 2016] and efficient pipelining on accelerators.

In practice, one of the key issues in ensuring an efficient execution of a general tensor contraction on a given device has been the inefficiency of the tensor transpose operation [Lyakh 2015], [Springer et al 2016a,b]. Since tensors are multidimensional objects, there exist a number of choices for the practical implementation of a tensor contraction. Hereafter we will call the two major algorithmic routes as *direct* and *indirect* approaches. In the indirect approach, a tensor contraction is formally transformed into the matrix-matrix multiplication (GEMM) via the tensor transpose operation, which may be required for each input/output tensor argument (up to four tensor transposes may be required in general). Here the efficiency of the tensor transpose operation is the main bottleneck as most vendor-provided GEMM routines are already highly optimized. Depending on a specific algorithm, the tensor transpose step needs to be performed either globally (HPC system level) [Solomonik et al 2013], [Rajbhandari et al 2014a,b] or only locally (node level), dictated by the data layout. Although the optimization of the global tensor transpose operation can be vital for the performance of relevant algorithms, here we will only consider local tensor transposes since the global tensor transpose step is not necessary in general. In contrast, in the direct approach [Baumgartner et al 2005], [Springer & Bientinesi 2016], [Matthews 2016], a tensor contraction is implemented explicitly as an optimized loop nest, without auxiliary data reshuffling, thus also avoiding the overhead associated with the temporary memory allocation required for the out-of-place tensor transpose. The advantages of the indirect approach are (a) it is fully general with the same (optimized) routine being applicable to any tensor contraction, (b) the vendor-provided GEMM routines are usually very efficient. On the other hand, in the direct approach, a separate optimized routine may be required for each individual tensor contraction case on each distinct computer architecture, thus making it an ideal target for automated code generation. In fact, few promising efforts have been reported very recently along this route [Springer & Bientinesi 2016], [Nielson et al 2015], [Matthews 2016]. Yet, the conceptual simplicity and generality of the indirect approach has stimulated a number of works towards optimization of the tensor transpose operation on CPU [Lyakh 2015], [Springer et al. 2016a], NVIDIA GPU [Lyakh 2015], [Springer et al. 2016b], and Intel Xeon Phi [Lyakh 2015], [Springer et al. 2016b]. Besides, the indirect approach is the only choice in cases where the required tensor contractions are not known at compile time [Lyakh & Bartlett 2010], [Lyakh 2014].

A generic tensor transpose algorithm has been presented in Ref. [Lyakh 2015], but the implementation of that algorithm was suboptimal in terms of performance, especially for NVIDIA GPUs. In the current work, we deliver a new optimized generic version of the tensor transpose operation for NVIDIA GPUs implemented as a standalone library under the name cuTT (CUDA Tensor Transpose). By carefully considering several limiting tensor transpose cases and introducing NVIDIA GPU specific optimizations, we are able to achieve an average performance 70-80% of the maximum achievable bandwidth on the NVIDIA Kepler, Maxwell, and Pascal GPU architectures. We compare cuTT to the auto-tuned CUDA code generation framework TTC introduced very recently for a similar purpose [Springer et al. 2016b], and show that the two produce close to identical performance. Finally, by integrating the cuTT library into the accelerated tensor algebra library TAL-SH, we demonstrate the efficiency of the new tensor transpose implementation by running a large sample of tensor contractions. Both the cuTT and TAL-SH libraries are publicly available on GitHub[1,2]. The main contributions of this work can be summarized as follows:

- We derive two main tensor transpose algorithms, called "Tiled" and "Packed" algorithms, that are both based on using shared memory buffers to reduce the scatter in global memory access patterns.
- We introduce an efficient thread-parallel algorithm for computing the global memory positions of tensor elements on GPU hardware, whose performance is independent of the tensor rank.
- We derive an analytical GPU performance model based on the existing MWP-CWP model that is used for picking the optimal tensor transpose algorithm and parameters at runtime without need for performance measurements or profiling.
- To the best of our knowledge, cuTT is the fastest public runtime library on NVIDIA GPUs for generic tensor transposes.
- We show that cuTT is as fast on NVIDIA GPUs as the TTC method that is based on code generation.
- We demonstrate that the tensor transpose overhead can be as low as few percent in compute intensive tensor contractions on NVIDIA GPUs.

## 2. ALGORITHMS

We will begin by going through some basic mathematical notation related to the tensor transpose operation. Let us assume we have a rank-$n$ tensor $T$, that is, a tensor with $n$ dimensions, $\{1, \ldots, n\}$, with extents $\{d(1), \ldots, d(n)\}$ and a total volume vol($T$) (i.e. number of elements in the tensor). In our notation, the first tensor dimension is the stride-1 dimension and the subsequent dimensions increase their seniority from left to right. Let us define the *cumulative volume* of the tensor dimensions for a permutation $\{w_j\}_1^n \equiv \{w_1, \ldots, w_n\}$ of indices $\{1, \ldots, n\}$ as

$$c\left(z, \{w_j\}_1^n\right) = \begin{cases} 1, & i = 1 \\ \prod_{j=1}^{i-1} d(w_j), & i > 1 \end{cases}$$

where

$$i = \arg_{j=1,\ldots,n}(z = w_j)$$

is the position of dimension $z$ in the permutation $\{w_j\}_1^n$. We define a *scalar position* in a rank-$n$ tensor as the linear position in the corresponding array in computer memory where the element $\{x_j\}_1^n \equiv \{x_1, \ldots, x_n\}, 0 \leq x_j < d(j)$ is stored

---

[1] https://github.com/ap-hynninen/cutt
[2] https://github.com/DmitryLyakh/TAL_SH

$$p\left(\{x_j\}_1^n, \{w_j\}_1^n\right) = \sum_{i=1}^n x_i c\left(w_i, \{w_j\}_1^n\right)$$

This map can be inverted to give the multi-linear coordinates $\{x_j\}_1^n$ back

$$x_i = \mod\left(\left\lfloor \frac{p(\{x_j\}_1^n, \{w_j\}_1^n)}{c(w_i, \{w_j\}_1^n)} \right\rfloor, d(w_i)\right),$$

where the integer division is rounded down and mod() denotes the integer modulo operation. We can now write the scalar position in the transposed tensor, corresponding to transposed coordinates $\{x_j\}_1^n$, based on the scalar position in the original tensor, corresponding to coordinates $\{y\}_1^n$, as

$$p\left(\{x_j\}_1^n, \{w_j\}_1^n\right) = \sum_{i=1}^n \mod\left(\left\lfloor \frac{p(\{y_j\}_1^n, \{j\}_1^n)}{c(i, \{j\}_1^n)} \right\rfloor, d(i)\right) c\left(w_i, \{w_j\}_1^n\right) \quad (1)$$

Equation (1) allows us to convert the scalar position $p\left(\{y_j\}_1^n, \{j\}_1^n\right)$ in the input tensor with the trivial index permutation $I = \{j\}_1^n$ into the scalar position $p\left(\{x_j\}_1^n, \{w_j\}_1^n\right)$ in the output tensor with the index permutation $O = \{w_j\}_1^n$. Using Equation (1), one can easily write a tensor transpose program that reads the input tensor elements in a linear fashion, $p\left(\{y_j\}_1^n, \{j\}_1^n\right) = 0, 1, \ldots, \text{vol}(T)-1$, and writes the elements into the output tensor positions given by the corresponding $p\left(\{x_j\}_1^n, \{w_j\}_1^n\right)$. Such a program implements a scattered tensor transpose algorithm. Since the output tensor positions are scattered, this algorithm is not expected to deliver good performance on modern computer architectures that rely on coalesced memory accesses.

We will briefly discuss the relevant features of the NVIDIA GPU SIMT architecture. Threads on NVIDIA GPUs are arranged into groups of threads called "thread blocks". Threads within a thread block are further divided into "warps" of 32 consecutive threads. Threads within a warp execute in lock step, similarly to vector operations on a CPU. The two parts of the GPU memory hierarchy that we need to consider for a high-performance tensor transpose algorithm are global memory and shared memory. Global memory is high-capacity off-chip memory shared among all threads. It has high access latency and can deliver 128 bytes in a single transaction on 128-byte aligned addresses. Shared memory is low-capacity on-chip memory shared among the threads in a thread block that has low access latency and allows for fast random accesses to individual memory elements, with the caveat of memory bank conflicts that limit performance on certain access patterns, see Ref. [NVIDIA 2015a].

Before describing the basic algorithm for tensor transposes on GPU, we need to define some nomenclature. The input and output tensors $I = \{1, \ldots, n\}$ and $O = \{w_j\}_1^n$ are partitioned into non-overlapping multi-indices $M_{mk}^I$ and $\bar{M}_{mk}^I$, and $M_{mk}^O$ and $\bar{M}_{mk}^O$, respectively. Multi-indices $M_{mk}^I$ and $M_{mk}^O$ consist of same indices but have different permutations: in $M_{mk}^I$ the indices are in input order while in $M_{mk}^O$ the indices are in output (i.e. transposed) order. Similarly, the indices in $\bar{M}_{mk}^I$ and $\bar{M}_{mk}^O$ are the same but in different order. We define the volume of the multi-index as the product of all the dimensions in the multi-index and drop the superscript when denoting volume since multi-indices with same indices but different permutation have the same volume, e.g. $\text{vol}(M_{mk}) = \text{vol}(M_{mk}^I) = \text{vol}(M_{mk}^O)$.

The multi-index $M_{mk}^I$ is further partitioned into two potentially overlapping multi-indices $M_m^I$ and $M_k^I$, where $M_m^I = \{1, \ldots, m\}$ consists of the first $m$ dimensions of the input tensor and $M_k^I$ consists of the first $k$ dimensions of the output tensor listed in the order they are in the input tensor. Similarly, multi-index $M_{mk}^O$ is partitioned into multi-indices $M_m^O$ and $M_k^O$, where $M_m^O$ is $M_m^I$ with output permutation of the $m$ dimensions and $M_k^O = \{w_1, \ldots, w_k\}$ is $M_k^I$ with output permutation of the $k$ dimensions. It is important observe that multi-indices $M_m^I$ and $M_k^O$ define contiguous sub-volumes of size $\text{vol}(M_m)$ and $\text{vol}(M_k)$, and therefore global memory reads and writes to these sub-volumes can be performed at high efficiency. This observation gives rise to the basic idea behind high-performance tensor transpose algorithms on GPU: Read $\text{vol}(M_{mk})$ using multi-index $M_{mk}^I$, which is contiguous up to width $\text{vol}(M_m)$, from global memory to shared memory. Then, by reading shared memory in transposed order, write $\text{vol}(M_{mk})$ using multi-index $M_{mk}^O$, which is contiguous up to width $\text{vol}(M_k)$, to global memory. The reason this algorithm performs well is that both the reads from global memory as well as writes to global memory are mostly coalesced; non-coalesced accesses are performed in shared memory when reading the shared memory in transposed order. This algorithm is illustrated in Figure 1: First sub-volume $\text{vol}(M_{mk})$ of the input tensor is read from global memory into shared memory. Then, by reading from the shared memory in a transposed order, the sub-volume $\text{vol}(M_{mk})$ is written into global memory of the output tensor. Note that although Figure 1 does not show this, the multi-indices $M_m^I$ and $M_k^I$ (and similarly $M_m^O$ and $M_k^O$) can overlap.

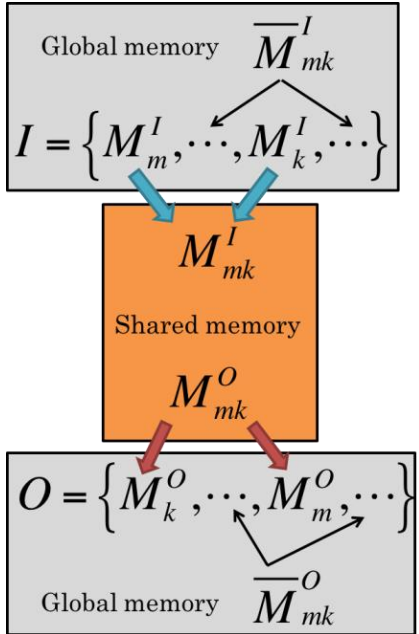

Fig. 1. Tensor transpose using a shared-memory. First, volume defined by the multi-index $M_{mk}^I$ is read from global memory into shared memory. Then, by reading the shared memory in a transposed order, the volume defined by the mult-index $M_{mk}^O$ is written to global memory.

The multi-indices $\bar{M}^I_{mk} \equiv \{1,...,n\} \setminus M^I_{mk}$ and $\bar{M}^O_{mk} \equiv \{w_j\}^n_1 \setminus M^O_{mk}$ determine the start reading and writing positions for the multi-indices $M^I_{mk}$ and $M^O_{mk}$, respectively. We refer to these reading and writing positions as "major" reading and writing positions. Let $\bar{M}^I_{mk} = \{s_1,...,s_h\}$, where the dimensions are in the input order and hence $s_1 < s_2 ... < s_h$. Using Equation (1), the major reading position from global memory is then obtained from

$$\text{pMajorIn}(b, \bar{M}^I_{mk}) = \sum_{i=1}^{h} \text{mod}\left(\left\lfloor \frac{b}{c(s_i, \bar{M}^I_{mk})} \right\rfloor, d(s_i)\right) c(s_i, I) \qquad (2)$$

where $b = [0,...,\text{vol}(\bar{M}_{mk}) - 1]$. Similarly, let $\bar{M}^O_{mk} = \{s_{l_1},...,s_{l_h}\}$, where the dimensions are in the output order given by $l_i \in [1,...,h]$. We can now obtain the major writing position to global memory from

$$\text{pMajorOut}(b, \bar{M}^O_{mk}) = \sum_{i=1}^{h} \text{mod}\left(\left\lfloor \frac{b}{c(s_{l_i}, \bar{M}^O_{mk})} \right\rfloor, d(s_{l_i})\right) c(s_{l_i}, O) \qquad (3)$$

where $b = [0,...,\text{vol}(\bar{M}_{mk}) - 1]$.

Our tensor transpose implementation relies on the efficient evaluation of Equations (2) and (3) on GPU hardware. In order to derive an algorithm to do this, it is important to note that parameter $b$ in Equations (2) and (3) is the same for all threads within the warp. We can therefore evaluate the elements of the sum in parallel on a warp and perform a parallel reduction to get the final result. If we restrict ourselves to $h \leq 32$ (which is enough for our purposes) we can perform the computation on a single warp, as shown in Algorithm 1.

---
**ALGORITHM 1.**  Tensor element position computation

**Given:** warpLane = 0, ..., L – 1 = Warp lane
```
int TensorPos(int b, int c, int d, int ct, int h)
{
    r = 0
    if warpLane < h then
        r = ((b / c) % d) * ct
    end
    unroll for (i = L / 2; i >= 1; i /= 2) do
        r += __shfl_xor(r, i)
    end
    return r
}
```
---

In Algorithm 1, we first evaluate the elements of the sum in Equation (1) in parallel on all warp lanes (which is thread index modulo L) less than $h$ and store the result in register r. It is important to note that the input variables c, d, and ct contain different values for every warp lane, while variables b and h are the same for all warp lanes. We then use the butterfly reduction [NVIDIA 2015a] to sum the elements together to get the final result in register r. The butterfly reduction uses the __shfl_xor operation that was introduced in the NVIDIA Kepler architecture, which allows threads within a warp to pass values to each other without using shared memory. Note that the result in r is present on all warp lanes.

Similarly to the major reading and writing positions in Equations (2) and (3), we now derive formulas for calculating the minor reading and writing positions in $M_{mk}^I$ and $M_{mk}^O$. Let $M_{mk}^I = \{q_1, ..., q_a\}$, where the dimensions are in the input order (hence $q_1 < q_2 ... < q_a$) and $a = \text{size}(M_{mk}^I)$ is the number of dimensions in $M_{mk}^I$. Further, let $M_{mk}^O = \{q_{t_1}, ..., q_{t_a}\}$, where the dimensions are in the output order given by $t_i \in [1, ..., a]$. We can now write the element positions as follows. The minor reading position from global memory is given by

$$\text{pMinorIn}(k, M_{mk}^I) = \sum_{i=1}^{a} \text{mod}\left(\left\lfloor \frac{k}{c(q_i, M_{mk}^I)} \right\rfloor, d(q_i)\right) c(q_i, I) \tag{4}$$

the minor writing position to global memory is given by

$$\text{pMinorOut}(k, M_{mk}^O) = \sum_{i=1}^{a} \text{mod}\left(\left\lfloor \frac{k}{c(q_{t_i}, M_{mk}^O)} \right\rfloor, d(q_{t_i})\right) c(q_{t_i}, O) \tag{5}$$

and finally the reading position from shared memory is given by

$$\text{pSh}(k, M_{mk}^O) = \sum_{i=1}^{a} \text{mod}\left(\left\lfloor \frac{k}{c(q_{t_i}, M_{mk}^O)} \right\rfloor, d(q_{t_i})\right) c(q_{t_i}, M_{mk}^I) \tag{6}$$

where $k = [0, ..., \text{vol}(M_{mk}) - 1]$.

Equations (2) – (6) provide the formulas required for calculation of element positions for the tensor transpose algorithm using a shared memory buffer. By utilizing Equations (2) – (6) and Algorithm 1, we will now develop the key algorithms that cuTT uses for tensor transposes.

### 2.1 Tiled Algorithm

Our tiled algorithm is a generalization of the basic 2D tiled matrix transpose algorithm [Harris 2013], where the 2D part is formed by the tensor dimensions in multi-indices $M_{mk}^I$ and $M_{mk}^O$, and we loop over the volume defined by the rest of the dimensions in multi-indices $\bar{M}_{mk}^I$ and $\bar{M}_{mk}^O$. In order to use this algorithm, the multi-indices $M_m^I$ and $M_k^O$ must each consist of a single dimension, in other words $M_m^I = M_m^O = \{1\}$ and $M_k^O = M_k^I = \{w_1\}$. If $M_m^I$ and $M_k^O$ have the same dimensions, i.e. $w_1 = 1$, we will be using a special "TiledCopy" algorithm where only the dimensions in $\bar{M}_{mk}^I$ and $\bar{M}_{mk}^O$ form a non-trivial permutation. The Tiled algorithm is illustrated in Figure 2 for a tile width $L=4$, $\text{vol}(M_m) = 9$, and $\text{vol}(M_k) = 7$, where the arrows show warps' memory access pattern: blue arrows indicate reading from memory and red arrows indicate writing to memory. In the first step shown in Figure 2(a), an $L \times L$ tile is read from global memory in dataIn and written to the shared memory buffer shBuffer. In the second step shown in Figure 2(b), the shared memory buffer shBuffer is read in a transposed order and written to global memory in dataOut. Finally, these two steps are repeated for every tile in $\text{vol}(M_m) \times \text{vol}(M_k)$ and for every element in $\text{vol}(\bar{M}_{mk})$. Note that the shared memory buffer shBuffer has size $L \times (L + 1) = 4 \times 5$. The reason for this is that had we used a $L \times L$ shared memory buffer, and assuming $L$ shared memory banks, there would be $L$ bank conflicts per read when the shared memory is read in transposed order in Figure 2(a).

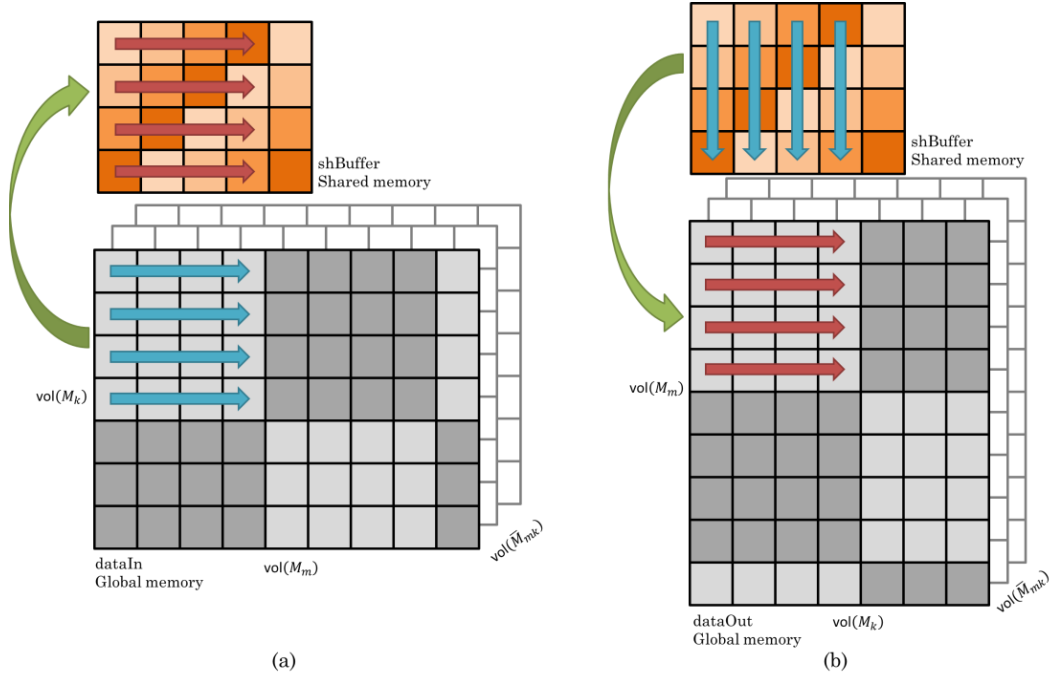

Fig. 2. Illustration of the Tiled algorithm for tile width $L=4$, $\text{vol}(\boldsymbol{M_m}) = 9$, and $\text{vol}(\boldsymbol{M_k}) = 7$, where in (a) global memory is read from dataIn and written to the shared memory buffer shBuffer, and in (b) the shared memory buffer shBuffer is read in a transposed order and written to global memory dataOut. Arrows show warps' memory access pattern where blue arrows indicate reading and red arrows indicate writing. The shades in shared memory buffer show the memory banks where each element belongs.

In the Tiled algorithm, the major element positions are calculated using Equations (2) and (3), and Algorithm 1. The minor element positions, given in Equations (4) – (6) simplify considerably as follows. Using $k = x + yd(1)$ with $0 \leq x < d(1)$ and $0 \leq y < d(w_1)$, in Equation (4) we obtain

$$\text{pMinorIn}(x, y, M_{mk}^I) = x + yc(w_1, I)$$

and using $k = y + xd(w_1)$ in Equation (5) we obtain

$$\text{pMinorOut}(x, y, M_{mk}^O) = y + xc(1, O)$$

Setting $d(1) = L + 1$, $d(w_1) = L$, and $k = t_x + t_y L$ with $t_y < L + 1$ and $t_x < L$ in Equation (6) we obtain

$$\text{pSh}(t_x, t_y, M_{mk}^O) = t_y + t_x(L + 1)$$

As mentioned above, we also implement a "TiledCopy" algorithm for the case where $w_1 = 1$. In the TiledCopy algorithm, there is no need for shared memory buffer since no transpose takes place within $M_{mk}$.

## 2.2 Packed Algorithm

The main limitation of the Tiled (and TiledCopy) algorithm is that $M_m^I$ and $M_k^O$ must each consist of a single dimension, in other words $M_m^I = \{1\}$ and $M_k^O = \{w_1\}$. This limits the performance of these algorithms to cases where the first input and output dimensions are large to allow for mostly coalesced memory accesses. Here we will present a "Packed" tensor transpose algorithm that allows us to pack multiple dimensions into both $M_m^I$ and $M_k^O$, and hence improve performance for the cases where the first input and output dimensions are small.

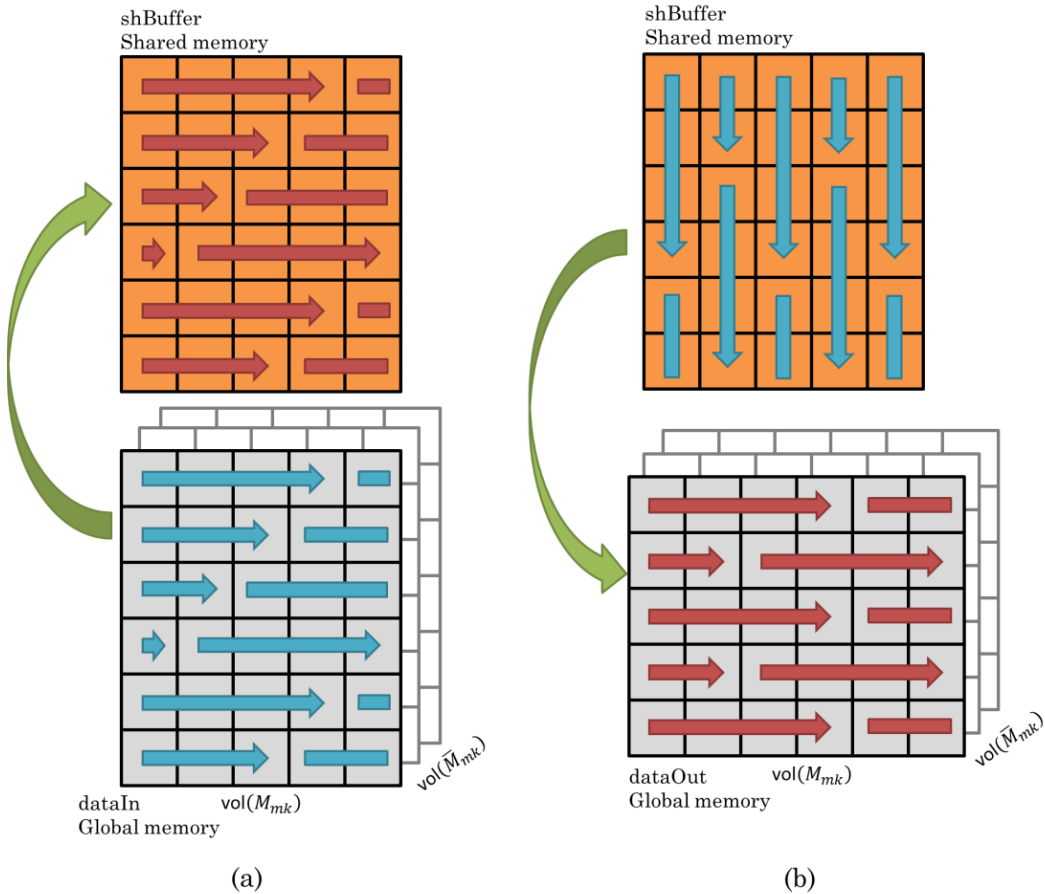

Fig. 3. Illustration of the Packed algorithm for $L=4$ and vol($\boldsymbol{M_{mk}}$) = **30**, where in (a) global memory is read from dataIn and written to the shared memory buffer shBuffer, and in (b) the shared memory buffer shBuffer is read in a transposed order and written to global memory dataOut. Arrows show warps' memory access pattern where blue arrows indicate reading and red arrows indicate writing.

Figure 3 illustrates the Packed algorithm for $L=4$ and vol($M_{mk}$) = 30, where the blue and red arrows indicate memory reading and writing, respectively. The basic steps of the algorithm are the same as in the Tiled algorithm: In the first step in Figure 3(a), global memory is read from dataIn and written into the shared memory buffer shBuffer. In the second step, in Figure 3(b), the shared memory buffer shBuffer is read in a transposed order and written to global memory in dataOut. The difference between the Tiled and the Packed algorithms is that, while in the Tiled algorithm exactly two dimensions are stored into the $L \times L$ shared memory space, in the Packed algorithm multiple dimensions are *packed* into the shared memory space of size $\text{vol}(M_{mk})$. The main limiting factor for using the Packed algorithm is therefore the size of the shared memory. At the end of this section, we will discuss how we get around this limitation by splitting one of the dimensions in $M_{mk}$.

In the Packed algorithm, the major global memory positions are calculated in the same way as in the Tiled algorithm, i.e. using Equations (2) and (3) and Algorithm 1. However, the minor memory positions given by Equations (4) - (6) must now be evaluated fully since these equations cannot be simplified in the general case. Moreover since variable $k$ in Equations (4) - (6) is not constant across the warp, we cannot use Algorithm 1 to evaluate these equations. However, since each thread accesses only a few distinct $k$ values, we can pre-calculate Equations (4) - (6) and store the results in registers. The size of the register arrays, $nReg$, is determined by $\text{vol}(M_{mk})$ and the number of threads per thread block, $nThread$, and is given by

$$nReg = \left\lceil \frac{\text{vol}(M_{mk})}{nThread} \right\rceil$$

Every thread now has its own values for pMinorOut, pMinorIn, and pSh that they can access very fast. The downside of this approach is increased register usage, which in turn can lead to lower occupancy and ultimately to register spilling. We limit the amount of register storage to $1 \leq nReg \leq 8$, for which no register spilling was observed.

As mentioned earlier in this section, use of the Packed algorithm is limited by the size of the shared memory, which must fit the entire volume $\text{vol}(M_{mk})$. In order to go around this limitation, we implemented a second version of the Packed algorithm, called the "PackedSplit" algorithm, where we split the largest dimension $g \in M_{mk}$ into $n_{sp}$ chunks such that the split volume $\text{vol}(M_{mk} \backslash g) \lceil d(g)/n_{sp} \rceil$ fits into shared memory. PackedSplit algorithm can handle efficiently many corner cases where both Tiled and Packed algorithms give low performance. An example of such a case is where the first input dimension is large and the first output dimension is small (less than warp-width $L$). In this case, Tiled algorithm will perform poorly because it will use the entire warp to perform only a few writes and Packed algorithm cannot be used at all since, due to large first input dimension, $\text{vol}(M_{mk})$ will not fit in shared memory.

### 2.3 Choosing the Optimal Tensor Transpose Plan

For many tensor transposes, it is possible to use multiple algorithms. Tiled (or TiledCopy if $w_1 = 1$) algorithm can always be used, while the Packed (or PackedSplit) algorithm can only be used when there exists a choice of $M_m^I$ and $M_k^O$ such that the $\text{vol}(M_{mk})$ (or $\text{vol}(M_{mk} \backslash g) \lceil d(g)/n_{sp} \rceil$) elements fit into GPU's shared memory. In the case of Tiled or TiledCopy algorithms, choice of multi-indices $M_m^I$ and $M_k^O$ is trivial. However, in the case of the Packed and PackedSplit algorithms, we must choose between different choices of multi-indices $M_m^I$ and $M_k^O$, and if the PackedSplit algorithm is used, we must additionally choose the number of splits $n_{sp}$. In summary, the algorithm and parameter choices are: Tiled (or TiledCopy if $w_1 = 1$), Packed with $\{M_m^I, M_k^O\}$, and PackedSplit with $\{M_m^I, M_k^O, n_{sp}\}$.

In cuTT the choice of the algorithm and the relevant parameters are packaged into a "plan". cuTT implements an API where a plan is first created, then executed, and finally destroyed, similarly to the popular libraries such as FFTW [Frigo & Johnson 2005] and cuFFT [NVIDIA 2015b]. This API structure enables multiple executions of the same tensor transpose without the overhead of initialization. Initialization involves memory allocations on the host and the GPU as well as copying of small memory arrays from the host to the GPU. Plan creation in cuTT takes as input the rank and dimension extents of the input tensor, as well as the permutation of the output tensor. At plan creation, cuTT creates all plans that can be executed on the target GPU. cuTT then determines the "best" plan and ignores the rest of the plans. A simple way to determine the best plan is to measure the runtime of tensor transpose execution for each plan and pick the fastest one. This measurement-based scheme is guaranteed to find the optimal plan, but it is only useful in cases where the same tensor transpose is executed multiple times and therefore the overhead of executing the measurement kernels can be absorbed. However, in many applications, such as the tensor contractions discussed in this manuscript, each tensor transpose is only executed once and therefore there is a need for a scheme that can pick the optimal algorithm with little overhead. To this end, we first tried simple "if-then-else" schemes based on some insight on how occupancy affects GPU performance as well as size of the vol($M_m$) and vol($M_k$) regions such as was done in [Lyakh 2015], but they were found to be unreliable in picking the optimal plan. We therefore turned our attention to a heuristics-based scheme where the number of clock cycles per plan execution is approximated based on a GPU performance model. We decided to use the MWP-CWP GPU performance model [Hong & Kim 2009], [Sim et al. 2012] due to its analytical form, accuracy, and because it can be used at runtime without profiling.

The MWP-CWP model is based on estimating the number of warps per streaming multiprocessor (SM) that can access memory simultaneously during one memory waiting period (MWP) and the number of warps that can finish their computation during the same period (CWP) [Hong & Kim 2009]. In our case the kernel execution is dominated by memory operations and we can therefore ignore the computation part, CWP. Applying the MWP-CWP model to the tensor transpose algorithms described in Sections 2.1 and 2.2, we can write the total number of execution cycles as

$$cycles = (cycles_{mem} + cycles_{shmem} + cycles_{ac})N_{iter} \qquad (10)$$

where $cycles_{mem}$ is the number of cycles taken by the global memory operations per "load global – store global" iteration, $cycles_{shmem}$ is the number of cycles taken by shared memory operations, $cycles_{ac}$ is the number of cycles taken by arithmetic and control operations, and $N_{iter}$ is the number of iterations. In Equation (10) we have left out the cycles spent on thread block synchronization since we observed these were not important for the model. Because our kernels are limited by the global memory bandwidth, it is reasonable to expect $cycles_{mem}$ to be the dominant term and we therefore spend most of our focus on it. In the MWP-CWP model, $cycles_{mem}$ is given by

$$cycles_{mem} = \frac{2 \times mem_{lat} \times MLP \times N_{warp}}{MWP} \qquad (11)$$

where factor 2 accounts for global memory loads and stores, $mem_{lat}$ is the average global memory latency, $MLP$ is the level of intra-warp memory level parallelism,

$N_{warp}$ is the number of warps per thread block, and MWP is the number of warps that can access memory simultaneously during one memory waiting period. In the MWP-CWP model, the average memory latency is given by

$$mem_{lat} = mem_{baselat} + (TPR_{mem} - 1)\Delta)$$

where $mem_{baselat}$ is the base memory access latency, $TPR_{mem}$ is the average number of transactions per memory request, and $\Delta$ is the departure delay of memory transaction. In the original MWP-CWP model [Hong & Kim 2009], the average number of transactions per memory request was given by

$$TPR_{mem} = \frac{LD_{tran} + ST_{tran}}{LD_{req} + ST_{req}} \qquad (12)$$

where $LD_{tran}$ and $ST_{tran}$ are the number of memory load and store transactions, and $LD_{req}$ and $ST_{req}$ are the number of memory load and store requests, respectively. The original MWP-CWP model was designed for pre-Kepler NVIDIA GPUs that did not have a unified L2 cache. The unified L2 cache was introduced in the NVIDIA Kepler architecture and is used in all post-Kepler architectures currently available. In NVIDIA GPUs with a unified L2 cache, all global memory loads and stores go through an L2 cache with 32-byte cache lines [NVIDIA 2015a]. While memory loads are still accurately described by the original MWP-CWP model in Equation (12) for GPUs with a unified L2 cache, memory stores are not. This is because memory stores that only access a portion of the 32-byte cache line will need to load that cache line from global memory first, then modify the elements in L2, and then write the cache line back to global memory. These partial stores to L2 cache lines therefore increase memory latency by a factor of two. On the other hand, writing full L2 cache lines has no extra memory latency penalty since there is no need for reading from global memory to fill the cache line before writing it back. Partial stores to L2 cache lines are taken into account in our new definition of the average number of memory transactions per memory request, which is given by

$$TPR_{mem} = \frac{LD_{tran} + ST_{tran}(1+CL_{part}/(CL_{full}+CL_{part}))}{LD_{req} + ST_{req}} \qquad (13)$$

where $CL_{part}$ and $CL_{full}$ are the number of partial and full L2 cache line stores, respectively. The factor $CL_{part}/(CL_{full} + CL_{part})$ in Equation (13) is the fraction of memory stores that are done to partial L2 cache lines.

$MLP$ in Equation (11) is the average number of independent memory requests a warp performs per iteration. For the Tiled algorithm, $MLP$ is calculated as an average over the number of full tiles ($T_{full}$), tiles cut in horizontal direction at memory read ($T_{horz}$), tiles cut in vertical direction in memory read ($T_{vert}$), and corner tiles that are cut in both directions ($T_{corn}$), and is given by

$$MLP = \frac{(L/R)(2T_{full} + T_{horz}) + \lceil v/R \rceil(T_{vert} + T_{corn}) + \lceil h/R \rceil(T_{horz} + T_{corn})}{2(T_{full} + T_{horz} + T_{vert} + T_{corn})}$$

where $L$ is the width of the warp, $R$ is the number of warps per thread block, $v = \mathrm{mod}(\mathrm{vol}(M_k), L)$ is the height of the tiles cut in the vertical direction on memory read and $h = \mathrm{mod}(\mathrm{vol}(M_m), L)$ is the width of the tiles cut in the vertical direction on memory read. For the TiledCopy algorithm the above formula simplifies to

$$MLP = \frac{(L/R)(T_{\mathrm{full}} + T_{\mathrm{horz}}) + \lceil v/R \rceil (T_{\mathrm{vert}} + T_{\mathrm{corn}})}{T_{\mathrm{full}} + T_{\mathrm{horz}} + T_{\mathrm{vert}} + T_{\mathrm{corn}}}$$

Note that for tensor transposes where $\mathrm{vol}(M_m)$ and $\mathrm{vol}(M_k)$ are multiples of $L$, $MLP = L/R$ for both Tiled and TiledCopy algorithms. For the Packed algorithm, we simply use $MLP$ equal to the level of register storage, $MLP = nReg$.

In the MWP-CWP model, parameter $MWP$ in Equation (11) is given by

$$MWP = \min(MWP_{mem} \times MLP, MWP_{peak}, N_{warpsPerSM})$$

where $N_{warpsPerSM}$ is the number of warps active per SM, $MWP_{mem} = mem_{lat}/\Delta$ is the number of warps that can access memory during one memory access period when the peak bandwidth is not reached, and

$$MWP_{peak} = \frac{mem_{BW}}{BW_{warp} \times N_{SM}}$$

is the number of warps that can access memory when the peak bandwidth is reached. Here, $mem_{BW}$ is the theoretical maximum memory bandwidth of the GPU, $N_{SM}$ is the number of SMs on the GPU and

$$BW_{warp} = \frac{Freq \times Bytes_{req}}{mem_{lat}}$$

is the bandwidth consumed by each warp, where $Freq$ is the GPU clock frequency and $Bytes_{req}$ is the number of bytes transferred through the memory bus per request, which we estimate using

$$Bytes_{req} = [TPR_{mem}(1 - cache_{hit}) + cache_{hit}] TranSize$$

where $cache_{hit}$ is the combined hit rate on L1 and L2 caches and $TranSize$ is the size of memory transactions (128 bytes for the GPUs used in this study). We use $cache_{hit} = 0.2$ in all cases. The number of shared memory cycles in Equation (10) is given by

$$cycles_{shmem} = 2 \times TPR_{shmem} \times shmem_{lat} \times MLP$$

where $TPR_{shmem}$ is the average number of shared memory transactions per shared memory request and $shmem_{lat}$ is the shared memory latency. For the Tiled algorithm, $TPR_{shmem} = 1$, since no bank conflicts are possible. For the Packed and PackedSplit algorithms, shared memory bank conflicts are possible on shared memory reads and therefore $TPR_{shmem}$ is calculated at runtime using the element positions given by Equation (6). Note that this calculation scales only up to the tensor volume in shared memory and can therefore be performed quickly.

The modified MWP-CWP model described above takes in parameters that depend on the GPU platform: $mem_{baselat}$, $\Delta$, $shmem_{lat}$, $cycles_{ac}$, $N_{SM}$, $mem_{BW}$, and $Freq$. Of these parameters $N_{SM}$, $mem_{BW}$, and $Freq$ are easily obtained by querying the GPU properties at runtime, while parameters $mem_{baselat}$, $\Delta$, and $shmem_{lat}$ are not publicly disclosed by NVIDIA. We obtain $mem_{baselat}$ and $\Delta$ by using a modified version of the P-Chase micro-benchmarking method [Mei et al. 2015] where instead of using a single thread, we employ 1 to 32 threads within a single warp each accessing memory from a different 128 byte section. Figure 4 shows the number of clock cycles per global memory access in the P-Chase kernel for 1 to 32 threads within a warp for Tesla K20X, Tesla M40, and Tesla P100 GPUs. Parameters $mem_{baselat}$ and $\Delta$ can be now determined by a linear fit of the data sets. As can be seen from Figure 4, the data for M40 and P100 GPUs is noisy compared to the data for K20X GPU. This is because for M40 and P100 GPUs we had to use clock counts that were measured as an average over the entire P-Chase kernel run, while for the K20X we were able to measure fine-grained instruction level clock counts [Mei et al. 2015].

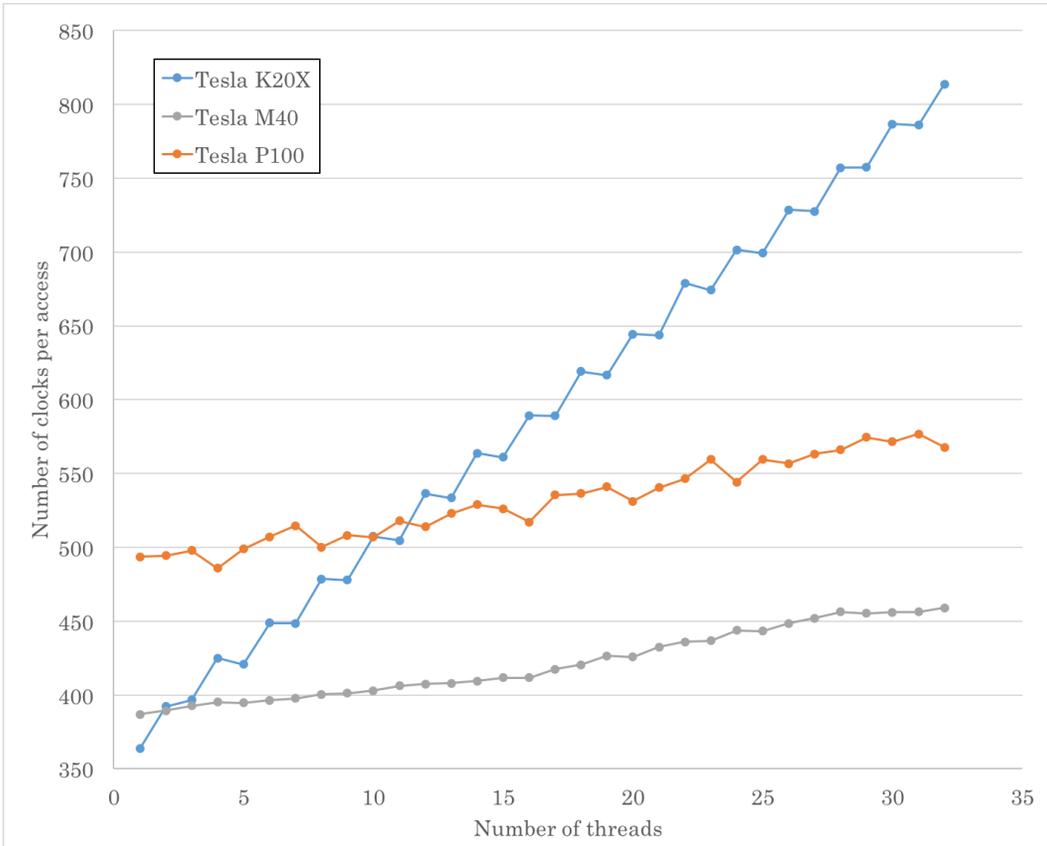

Fig. 4. Number of clock cycles per global memory access in the fine-grained P-Chase kernel for 1 to 32 threads on a single warp on Tesla K20X, Tesla M40, and Tesla P100 GPUs.

The rest of the parameters, $shmem_{lat}$ and $cycles_{ac}$ were used as a fitting parameters to our MWP-CWP model and were optimized to minimize the number of mispredictions the heuristic scheme makes in choosing the optimal plan compared to the measurement-based scheme. The resulting parameter values are summarized in Table I.

**Table I. Heuristic model parameters**

|  | Δ | $mem_{baselat}$ | $shmem_{lat}$ | $cycles_{ac}$ |
|---|---|---|---|---|
| Kepler (SM 3.X) | 14 | 358 | 11 | 50 |
| Maxwell (SM 5.X) | 2.5 | 385 | 1 | 220 |
| Pascal (SM 6.X) | 2.8 | 485 | 1 | 260 |

In order to use our MWP-CWP model, we need to calculate the average number of transactions per memory request, $TPR_{mem}$, given by Equation (13). This in turn requires us to know the number of memory requests ($LD_{req}$ and $ST_{req}$), memory transactions ($LD_{tran}$ and $ST_{tran}$), and the number of partial and full L2 stores ($CL_{part}$ and $CL_{full}$). While the number of memory requests ($LD_{req}$ and $ST_{req}$) can be easily calculated using an analytical formula, the same is not true for the number of memory transactions ($LD_{tran}$ and $ST_{tran}$) and for the number of L2 stores ($CL_{part}$ and $CL_{full}$). This is because their values depend on the memory alignment properties of the tensor transpose and can only be evaluated exactly by evaluating every element of the tensor. Such a computation would be too costly, and we have therefore resorted to a statistical estimation of $LD_{tran}$, $ST_{tran}$, $CL_{part}$, and $CL_{full}$, where these parameters are calculated for a set of ten random values from vol($\bar{M}_{mk}$). Cost of each of these calculations is limited by vol($M_{mk}$) and hence by the size of shared memory, making this estimation tractable. We have compared the statistical estimate to the exact calculation and found that it does not produce an important additional source of error to the heuristic scheme.

Our MWP-CWP model is clearly a simplification. Among many approximations, we believe the biggest one is the use of a fixed cache hit rate $cache_{hit}$ for all transposes, instead of an actual estimate based on kernel and transpose properties. However, as we show in the Results Section, our model is a useful approximation that can be used to choose the optimal algorithm for tensor transposes at runtime without incurring much overhead.

### 3. RESULTS

We perform benchmark bandwidth measurements on four NVIDIA GPUs: Tesla K20X, Tesla K40m, Tesla M40, and Tesla P100. On all GPUs the ECC is on and K40m is running in the non-boost mode. The code was compiled using CUDA 7.5 for Tesla K20X and K40m GPUs, and CUDA 8.0 for Tesla M40 and P100 GPUs. We had L1 caching switched on at compile time using the "-Xptxas -dlcm=ca" CUDA compiler flag and we used level 3 optimization ("–O3") throughout.

For each GPU, we determined the maximum attainable memory bandwidth using the GPU-STREAM program [Deakin et al. 2016] that was compiled and run on the same hardware and software setups as the cuTT benchmarks. Benchmarks for Kepler (Tesla K20X and K40m) and Pascal (Tesla P100) architectures are performed on tensors with 8-byte (double precision) elements, while those for Maxwell (Tesla M40) architecture are performed with 4-byte (single precision) elements. We benchmark Maxwell architecture using 4-byte tensor elements because we expect that most workloads on Maxwell GPUs are going to be done in single precision due to the low double-precision floating-point operation throughput. The maximum memory bandwidths reported by GPU-STREAM for 200M 8-byte elements were: 179 GB/s for Tesla K20X, 190 GB/s for Tesla K40m, 224 GB/s for Tesla M40, and 541 GB/s for

Tesla P100. In what follows, these values are used for reporting "Percentage of the maximum bandwidth" results.

### 3.1 Benchmark Set 1

The first set of benchmarks has the following characteristics.

(1) Tensor ranks 2 to 7.
(2) Tensor volume is normally distributed with an average of 200M and standard deviation of 40M elements.
(3) Ratio between the largest and the smallest tensor dimension is 1:1, 5:1, or 15:1.
(4) 500 random permutations for each tensor rank.

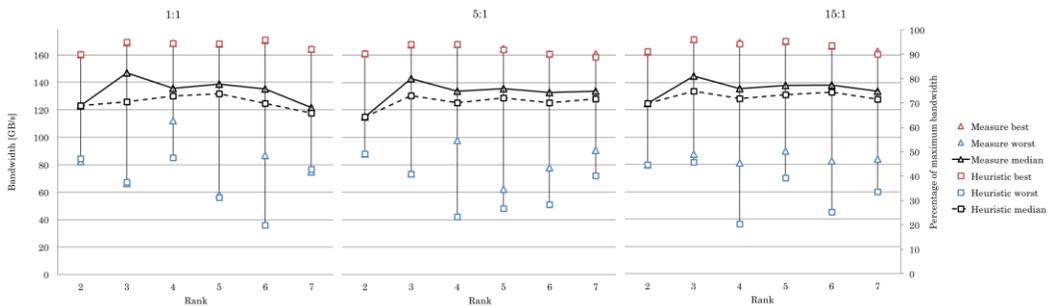

Fig. 5. Benchmark results on Tesla K20X for largest-to-smallest tensor dimension ratios 1:1, 5:1, and 15:1. Tensor element size is 8 bytes.

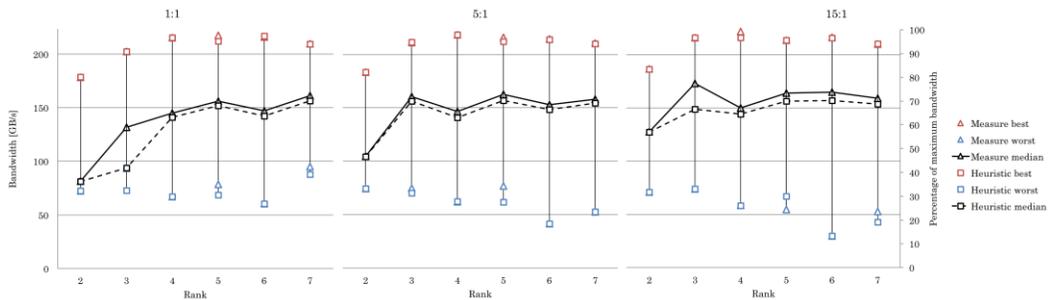

Fig. 6. Benchmark results on Tesla M40 for largest-to-smallest tensor dimension ratios 1:1, 5:1, and 15:1. Tensor element size is 4 bytes.

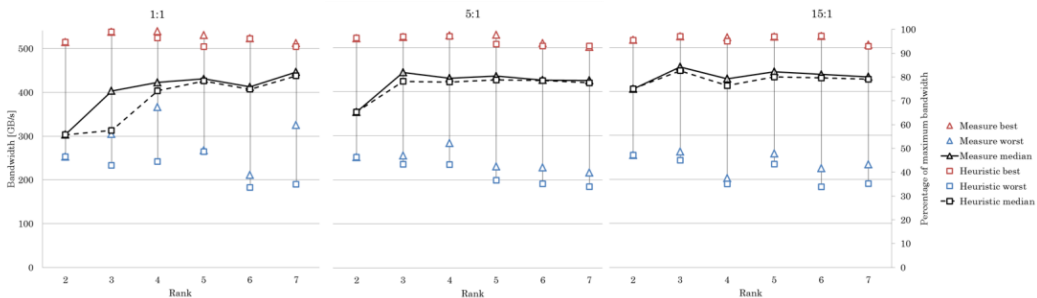

Fig. 7. Benchmark results on Tesla P100 for largest-to-smallest tensor dimension ratios 1:1, 5:1, and 15:1. Tensor element size is 8 bytes.

In Figures 5, 6, and 7, we show the benchmark results for Tesla K20X, Tesla M40, and Tesla P100 GPUs, respectively. Bandwidths are calculated using

$$\text{Bandwidth} = 2\frac{\text{vol}(T)}{D}E$$

where vol($T$) is the number of tensor elements (tensor volume), $E$ is the size of the elements in bytes, and $D$ is the CUDA kernel execution time in seconds. The results in Figures 5-7 show the worst, median, and best bandwidths for both heuristic and measurement based plans. As expected, the measurement-based plans always outperform the heuristic plans. The median bandwidths for Tesla K20X, shown in Figure 5, are around 70-80% of the maximum bandwidth for both heuristic and measure based plans. For Tesla M40, shown in Figure 6, the median bandwidths hover around 70% of the maximum bandwidth in most cases. For Tesla P100, shown in Figure 7, the median bandwidth is around 80% of the maximum bandwidth in most cases and the worst bandwidths are around 40% of the maximum bandwidth.

Notably, suboptimal performance is seen for rank-2 tensors with 1:1 ratio on Tesla M40 GPU in Figure 6 and Tesla P100 GPU in Figure 7, but not on Tesla K20X GPUs in Figure 5. In order to investigate this phenomenon further, we performed a series of transposes for single-precision rank-2 tensors with the 1:1 ratio, i.e. square matrices, with side dimension ranging from 13952 to 13968. Figure 8 shows the bandwidths we obtained as a percentage of the maximum bandwidth for Tesla K20X, M40, and P100 GPUs. The horizontal axis in Figure 8, modulo 32 of the matrix side dimension, measures the alignment of the memory accesses: At mod(dimension,32)=0 all memory accesses are perfectly aligned to 128 bytes. Square matrix transposes are a special case where, with suitable choice of dimensions, practically all memory accesses can become unaligned. From Figure 8 we see that for square matrices, 14 out of 16 choices for dimensions give rise to unaligned memory accesses. Therefore, if the dimensions are picked at random, like they are in our benchmark, it is highly likely to pick dimensions that give rise to many unaligned memory accesses. We see from Figure 8 that while the Tesla M40 GPU has high bandwidth for perfectly aligned data, performance degrades more for misaligned data than it does for Tesla K20X and Tesla P100 GPUs. Tesla K20X has the best performance overall in terms of percentage of the maximum bandwidth, while Tesla P100 is somewhere in the middle. This is the reason we see poor performance for 1:1 ratio rank-2 tensors in Figures 6 and 7, but not in 5. This analysis also suggests that it is a good idea to use at least 32 byte (i.e. 8 single-precision elements with mod(dimension,32)=8) aligned dimensions for square matrix transposes.

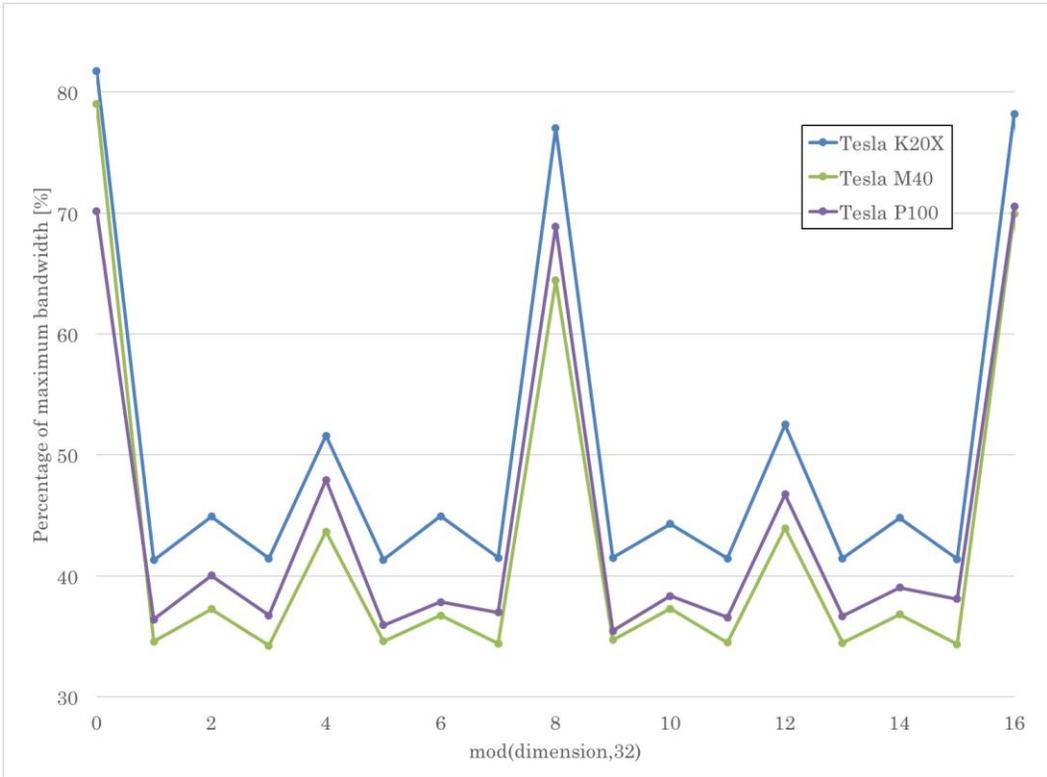

Fig. 8. The square matrix transpose bandwidth as a percentage of the maximum bandwidth for side dimension 13952 to 13968 of 4-byte elements for Tesla K20X, M40, and P100 GPUs. The horizontal axis plots modulo 32 of the matrix dimension.

### 3.2 Benchmark Set 2

In the second set of benchmarks we have tensors of rank 8 and 12, each having four large dimensions and all others small. Specifically, the dimensions for the rank-8 and rank-12 tensors are {5, 3, 2, 4, 35, 33, 37, 40} and {2, 3, 4, 3, 2, 2, 3, 2, 20, 18, 22, 24}, respectively. The rank-8 tensor has approximately 200M elements and the rank-12 tensor has approximately 328M elements. For both tensors, we evaluated 500 random permutations, as well as the trivial and reverse permutations.

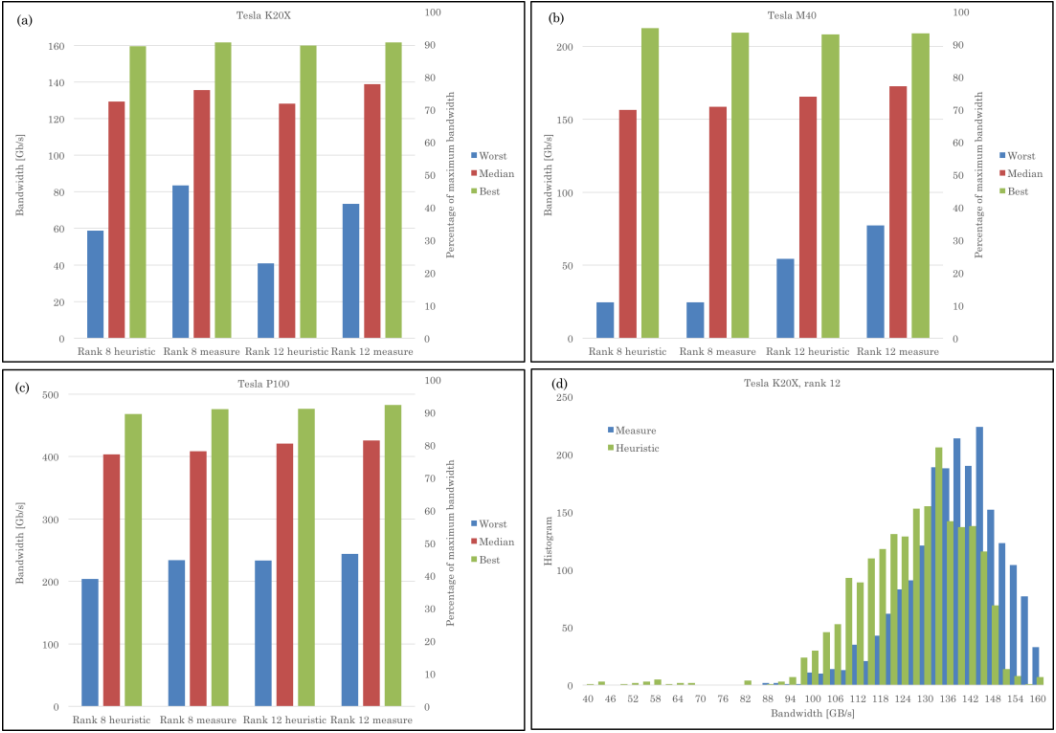

Fig. 9. Bandwidth results for rank-8 and rank-12 tensors on (a) Tesla K20X, (b) Tesla M40, (c) Tesla P100 GPUs. The distribution statistics is shown in (d) for Tesla K20X.

Benchmark results are shown in Figures 9(a)-(c) for Tesla K20X, M40, and P100 GPUs, where we plot the worst, best, and median bandwidths for both heuristics and measurement-based tensor transpose plans. The median bandwidths are around 70-80% of the maximum bandwidth, but the distribution of bandwidths is very wide, from 10% to 90% of the maximum bandwidth. Due to this large variation in bandwidths, it is interesting to look at how they are distributed. This is shown in Figure 9(d) that plots the histograms of bandwidths for Tesla K20X for rank-12 tensors. As can be seen from Figure 9(d), the heuristics-based tensor transpose plans have a long tail on the low bandwidth side that gives rise to the worst cases recorded in Figure 9(a), while the median bandwidth is not affected as much.

### 3.3 Benchmark Set 3

The final set of tensor transpose benchmarks consists of the 57 tensor transposes that were used in evaluating the performance of the TTC compiler by Springer et al. [Springer et al. 2016b], thus enabling us to make a direct comparison with their approach. In order to facilitate comparison with their results, we modified our CUDA kernels such that instead of just writing to global memory, we perform accumulation exactly as was done by Springer et al.: read input, read output, accumulate, write output. Since there are three global memory operations now, the bandwidth is calculated as

$$\text{Bandwidth} = 3 \frac{\text{vol}(T)}{D} E$$

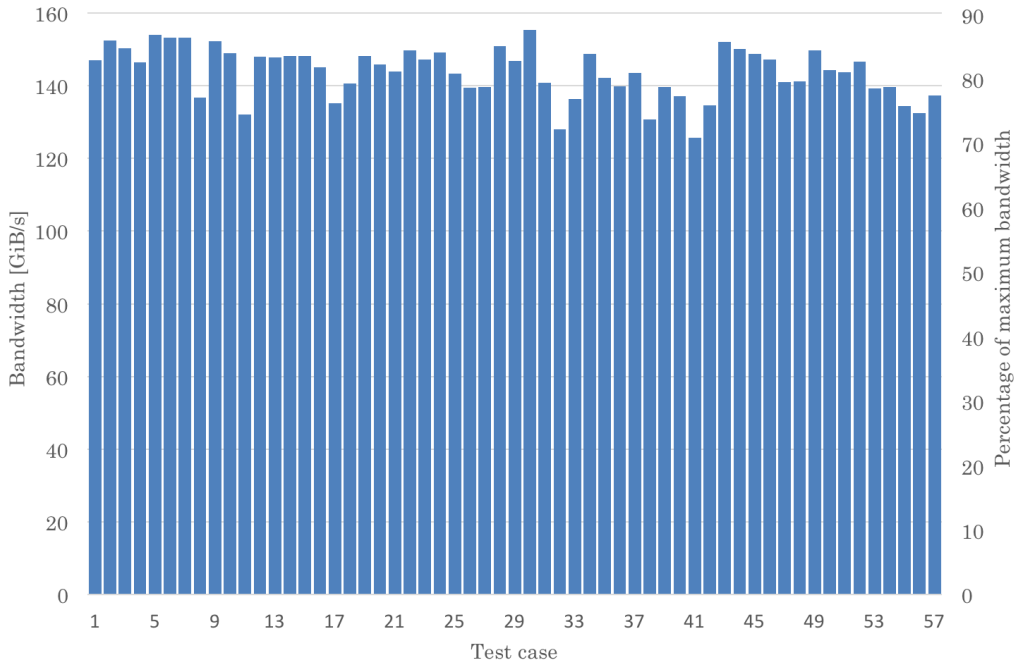

Fig. 10. Benchmarks for 57 tensor transposes used by Springer et al. [Springer et al. 2016b] on Tesla K40m.

Our benchmark results are shown in Figure 10 where we measure the bandwidth in GigaBytes (GiB) per second in order to facilitate direct comparison with Springer et al. [Springer et al. 2016b]. Comparing our results in Figure 10 with those by Springer et al., we see that they are quite similar: both hover in the range 120 to 160 GiB/s. Our benchmarks are slightly better in that they never dip below 120 GiB/s while those by Springer et al. do. However, overall both benchmarks can be said to show essentially the same performance. The median of the results shown in Figure 9 is 145 GiB/s or 82% of the maximum bandwidth.

**3.4. Tensor Contraction Benchmarks**
In addition to the pure tensor transpose benchmarks in Sections 3.1-3.3, we also benchmarked a large sample of random tensor contractions, in which the overall performance and the tensor transpose overhead were measured. As we mentioned in the introduction, in the indirect implementation of the tensor contraction operation the performance of the tensor transpose step is crucial for efficiency [Lyakh 2015]. In general, a binary tensor contraction can involve up to four tensor transposes (two forward transposes for the two input tensors, one forward and one backward transpose for the output tensor). Clearly, for memory-bandwidth-bound tensor contractions (tensor contractions with low arithmetic intensity), the overhead of the tensor transpose steps will necessarily be significant, but at least not as high as it could be without using the optimized tensor transpose algorithm presented in this paper. For compute-bound tensor contractions (tensor contractions with high arithmetic intensity), we further decrease the tensor transpose overhead, achieving as low as 1% (or even less) for some highly arithmetically intensive tensors contractions.

The benchmarks presented below were performed with the use of the tensor algebra library TAL-SH which implements basic tensor algebra operations on multi-core CPU and NVIDIA GPU, supporting multiple GPU accelerators per node. The cuTT library was integrated into TAL-SH to ensure efficiency of the tensor transpose step in tensor contractions. In the Tesla K20X benchmarks, both codes were compiled with the GNU compiler version 5.3.0 and NVIDIA CUDA 7.5. In the Tesla M40 and P100 benchmarks, both codes were compiled with the GNU compiler version 5.4.0 and NVIDIA CUDA 8.0. The Tesla K20x benchmarks were run on a single node of the Titan supercomputer at the Oak Ridge Leadership Computing Facility. The Tesla M40 and P100 benchmarks were run on the PSG development cluster at NVIDIA.

Our benchmark sample comprises 9306 random tensor contractions, where all possible tensor contraction configurations involving tensors up to rank 8 were considered. Also, each tensor contraction was run twice. For a given tensor rank, we randomly choose the location of the contracted and uncontracted dimensions while limiting ourselves up to 64 specific tensor contraction patterns for each unique tensor contraction configuration (in some cases the underlying tensor contraction configuration contained less than 64 unique patterns). In each input tensor, up to 4 large (unbounded) dimensions were allowed; the extent of other dimensions was restricted to 8. The large dimensions were chosen as the leading ones in the tensor storage layout. In each tensor contraction on K20X, the volume of the largest tensor was adjusted from below to be as close as possible to 75M elements of double precision or 150M elements of single precision, which was enforced by scaling up its large (unbounded) dimensions (also causing the change in the corresponding dimensions in other tensors). In the M40 and P100 benchmarks, the corresponding target volumes were doubled to 150M and 300M elements of double and single precision, respectively. High-rank tensors could require downscaling of all their dimensions to fit within the specified tensor volume limit (clearly, higher tensor ranks result in smaller tensor dimensions). It is important to note that all benchmarked tensor contractions required at least one tensor transpose step, and many of them required more than 1, with up to 4 in general.

Figures 11(a) and 11(b) show the GFlop/s (billion floating-point operations per second) our benchmarks achieved on Tesla K20X in single and double precision, respectively. The time a tensor contraction took was measured using CUDA events, starting from the first tensor transpose kernel on GPU and ending right after the last operation on GPU. The floating-point performance is plotted against arithmetic intensity of tensor contractions (the tensor contraction sample was binned into 8 subranges with respect to arithmetic intensity). Arithmetic intensity for tensor contraction $D = D + L \cdot R$ is given by

$$AI = \frac{2\sqrt{\text{vol}(D)\text{vol}(L)\text{vol}(R)}}{\text{vol}(D) + \text{vol}(L) + \text{vol}(R)}$$

where $\text{vol}(D)$, $\text{vol}(L)$, and $\text{vol}(R)$ are the tensor volumes (i.e., number of elements). As one can see, the plots have a similar shape and the single-precision performance is about twice higher than the double-precision performance. Arithmetic intensities of few thousands recover most of the Flop/s delivered by the optimized cuBLAS GEMM routine, that is, the tensor transpose overhead is negligible (on average). However, the performance drops quickly with decreasing arithmetic intensity (moving from the compute-bound to memory-bandwidth-bound region). Tensor contractions with arithmetic intensity less than a thousand are less likely candidates for GPU accelerated computing.

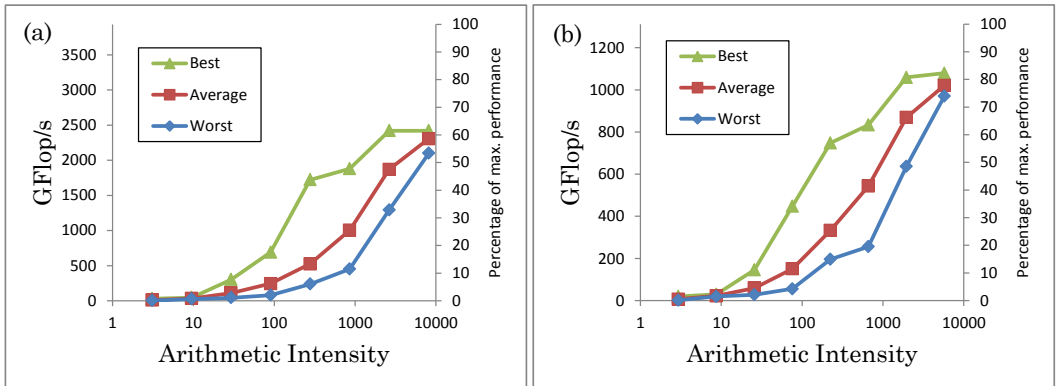

Fig. 11: Tensor contraction performance on Tesla K20X: (a) Single precision; (b) Double precision. The performance is measured against arithmetic intensity of tensor contractions.

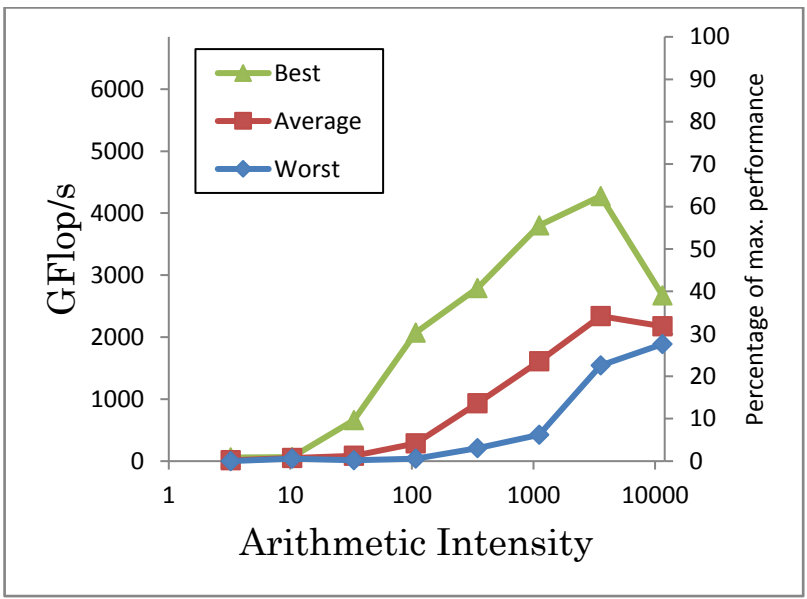

Fig. 12: Tensor contraction performance on Tesla M40: Single precision. The performance is measured against arithmetic intensity of tensor contractions.

Figure 12 shows the single-precision only performance on Tesla M40[3]. We can notice a decline in the average performance as compared to the best-case performance chart. Also, for some reason, we observed a drop in the cuBLAS SGEMM performance for very high arithmetic intensities, causing a noticeable drop in the best-case tensor contraction performance that can be seen in the far right part of the plot. Finally, Figure 13(a) and 13(b) illustrate the single- and double-precision performance on Tesla P100, achieving an impressive ~8 TFlop/s and ~4.5 TFlop/s at best, respectively. Also, on Tesla P100, the average double-precision performance is noticeably closer to the best-case performance, making it a very attractive GPU choice for numeric tensor algebra workloads.

---

[3] Due to resource usage limits, we could only benchmark about 90% of the full tensor contraction sample on NVIDIA M40. However, this is not expected to noticeably affect the performance results.

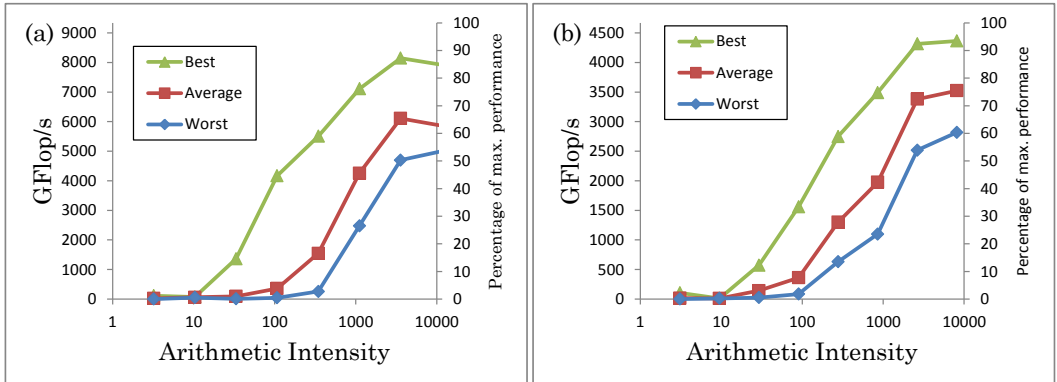

Fig. 13: Tensor contraction performance on Tesla P100: (a) Single precision; (b) Double precision. The performance is measured against arithmetic intensity of tensor contractions.

### 4. CONCLUSIONS

We have presented a high-performance tensor transpose library for CUDA compatible GPUs that provides an efficient implementation of the general tensor transpose operation independent of the tensor rank.

At its core, the library consists of two algorithms, called the Tiled and the Packed algorithms, that are both based on maximizing the amount of coalesced global memory accesses by using a shared memory buffer, but otherwise differ in their memory access patterns. The key innovation that enables us to achieve high performance independent of the tensor rank is the warp-parallel calculation of global memory element positions presented in Algorithm 1.

By performing extensive benchmarks on three generations of NVIDIA GPUs, we show that cuTT on average achieves approximately 70-80% of the maximum attainable bandwidth for Kepler, Maxwell, and Pascal GPU architectures. Our benchmarks include tensor ranks 2 up to 12 and show that the cuTT library performance is independent of the tensor rank. We also make comparison with the performance of the TTC compiler by Springer et al. [Springer et al. 2016b] and show that cuTT, a fully runtime library, achieves the same performance as this compiler based approach. We believe further improvements to cuTT performance are possible. To this end, it is useful to look for tensor transpose examples that have the worst performance and try optimizing those.

We derived and implemented an advanced heuristic scheme for picking the optimal plan for tensor transposes that is based on the MWP-CWP GPU performance model. Our heuristic scheme is able to pick the optimal plan for tensor transposes reliably: Benchmarks using heuristic plans are competitive against benchmarks using measurement-based plans. However, improvements to the heuristic scheme are welcome since it suffers, in addition to not being 100% accurate, from other inadequacies: With the introduction of new GPU architectures, the parameters have to be re-fitted, or worse, the entire performance model has to be redefined. Also, the computation of the number of memory accesses that is required for the heuristic scheme can become CPU compute intensive for high rank tensors especially on CPUs that have relatively low single core performance such as the IBM Power CPUs. Alternatives to using the MWP-CWP performance model could include approaches that use decision trees or machine learning.

By integrating the cuTT library into the tensor algebra library TAL-SH, we have shown that cuTT can decrease the tensor transpose overhead in the indirect GPU implementation of the tensor contraction operation down to 1% (or even less) for high arithmetic intensity tensor contractions. Yet, the tensor transpose overhead is growing very quickly with decreasing arithmetic intensity. To this end, possibly specialized approaches based on fusing the matrix multiplication and transpose kernels could yield better performance.

The work presented in this article specifically focused on the efficient implementation of tensor transposes on GPU and it is not clear how well the same algorithms could be adapted to CPU platforms. However, the idea presented in Algorithm 1 of calculating the tensor element positions in parallel should be applicable to CPUs with wide vectors, particularly the Intel Xeon Phi processors.

ACKNOWLEDGMENTS

*This research used resources of the Oak Ridge Leadership Computing Facility, which is a DOE Office of Science User Facility supported under Contract DE-AC05-00OR22725. We would like to thank NVIDIA, and specifically Mark Berger, for giving us access to Tesla M40 and Tesla P100 GPUs on their PSG cluster. DIL would like to thank Frank Winkler for his help in performance analysis.*

REFERENCES

Woody Austin, Grey Ballard, and Tamara G. Kolda. 2016. Parallel tensor compression for large-scale scientific data. *arXiv*:1510.06689 (2016).

Ning Bao, ChunJun Cao, Sean M. Carroll, Aidan Chatwin-Davies, Nicholas Hunter-Jones, Jason Pollack, and Grant N. Remmen. 2015. Consistency conditions for an AdS/MERA correspondence. *arXiv*:1504.06632 (2015).

Rodney J. Bartlett and G. D. Purvis. 1980. Molecular applications of coupled-cluster and many-body perturbation methods. *Physica Scripta* 21, (1980), 255-265. DOI: 10.1088/0031-8949/21/3-4/007.

Rodney J. Bartlett and Monica Musial. 2007. Coupled-cluster theory in quantum chemistry. *Reviews of Modern Physics* 79, (2007), 291-352. DOI:10.1103/RevModPhys.79.291.

G. Baumgartner, A. Auer, D. E. Bernholdt, A. Bibireata, V. Choppella, D. Cociorva, X. Gao, R. J. Harrison, S. Hirata, S. Krishnamoorthy, and others. 2005. Synthesis of high-performance parallel programs for a class of ab initio quantum chemistry models. *Proc. IEEE 93, 276-292 (2005)*.

J. A. Calvin, C. A. Lewis, E. F. Valeev. 2015. Scalable task-based algorithm for multiplication of block-rank-sparse matrices. *IA3 '15 Proceedings of the 5th Workshop on Irregular Applications: Architectures and Algorithms,* Austin TX, Nov 15, 2015, ISBN: 978-1-4503-4001-4.

Garnet Kin-Lic Chan, Anna Keselman, Naoki Nakatani, Zhendong Li, Steven R. White. 2016. Matrix product operators, matrix product states, and ab initio density matrix renormalization group algorithms. *Journal of Chemical Physics* 145, (2016), 014102. DOI: 10.1063/1.4955108.

Tom Deakin, James Price, Matt J. Martineau M, and Simon N. McIntosh-Smith. 2016. GPU-STREAM v2.0: Benchmarking the achievable memory bandwidth of many-core processors across diverse parallel programming models. 2016. Paper presented at P^3MA Workshop at ISC High Performance, Frankfurt, Germany.

Evgeny Epifanosky, Michael Wormit, Tomasz Kus, Arie Landau, Dmitry Zuev, Kirill Khistyaev, Prashant Manohar, Ilya Kaliman, Andreas Dreuw, and Anna I. Krylov. 2013. New implementation of high-level correlated methods using a general block tensor library for high-performance electronic structure calculations. *Journal of Computational Chemistry* 34, (2013), 2293-2309.

Matteo Frigo and Steven G. Johnson. 2005. The Design and Implementation of FFTW3. *Proceedings of the IEEE* 93, 2 (2005), 216-231. `DOI`:http://dx.doi.org/10.1109/JPROC.2004.840301

Peter Goldsborough. 2016. A tour of TensorFlow. *arXiv*:1610:01178 (2016).

Gaute Hagen, Thomas Papenbrock, M. Hjorth-Jensen, and David J. Dean. 2014. Coupled-cluster computations of atomic nuclei. *Reports on Progress in Physics* 77, (2014), 096302. DOI: 10.1088/0034-4885/77/9/096302.

Mark Harris. 2013. An Efficient Matrix Transpose in CUDA C/C++. (2013). In *Parallel Forall* Blog, Retrieved July 19, 2016 from https://devblogs.nvidia.com/parallelforall/efficient-matrix-transpose-cuda-cc/

Sunpyo Hong and Hyesoon Kim. 2009. An analytical model for a GPU architecture with memory-level and thread-level parallelism awareness. SIGARCH Comput. Archit. News 37, 3 (June 2009), 152-163. DOI=http://dx.doi.org/10.1145/1555815.1555775

Dmitry I. Lyakh and Rodney J. Bartlett. 2010. An adaptive coupled-cluster theory: @CC approach. *Journal of Chemical Physics* 133, (2010), 244112.

Dmitry I. Lyakh, Monica Musial, Victor F. Lotrich, and Rodney J. Bartlett. 2012. Multireference nature of chemistry: The coupled-cluster view. *Chemical Reviews* 112, (2012), 182-243. DOI: 10.1021/cr2001417.

Dmitry I. Lyakh and Rodney J. Bartlett. 2014. Algebraic connectivity analysis in molecular electronic structure theory II: Total exponential formulation of second-quantized correlated methods. *Molecular Physics* 112, (2014), 213-260. DOI: 10.1080/00268976.2013.807946.

Dmitry I. Lyakh. 2014. Scale-adaptive tensor algebra for local many-body methods of electronic structure theory. *International Journal of Quantum Chemistry* 114, (2014), 1607-1618. DOI: 10.1002/qua.24732.

Dmitry I. Lyakh. 2015. An efficient tensor transpose algorithm for multicore CPU, Intel Xeon Phi, and NVidia Tesla GPU. *Computer Physics Communications* 189, (2015), 84-91. DOI:http://dx.doi.org/10.1016/j.cpc.2014.12.013

D. A. Matthews. 2016. High-performance tensor contraction without BLAS. *arXiv*:1607.00291 (2016).

Xinxin Mei and Xiaowen Chu. 2015. Dissecting GPU Memory Hierarchy through Microbenchmarking. In *IEEE Transactions on Parallel and Distributed Systems*. September 2015. DOI: 10.1109/TPDS.2016.2549523

Naoki Nakatani and Garnet Kin-Lic Chan. 2013. Efficient tree tensor network states (TTNS) for quantum chemistry: Generalizations of the density matrix renormalization group algorithm. *Journal of Chemical Physics* 138, (2013), 134113. DOI: 10.1063/1.4798639.

T. Nielson, A. Rivera, P. Balaprakash, M. Hall, P. D. Hovland, E. Jessup, and B. Norris. 2015. Generating efficient tensor contractions for GPUs. *ICPP'15 Proceedings of the 2015 44th International Conference on Parallel Processing,* (2015), P. 969-978.

NVIDIA. 2015a. CUDA C Programming Guide version 7.5

NVIDIA. 2015b. CUFFT Library User's Guide version 7.5

S. Rajbhandari, A. Nikam, P.-W. Lai, K. Stock, S. Krishnamoorthy, and P. Sadayappan. 2014. A communication-optimal framework for contracting distributed tensors. *Proceedings of the SC14: International Conference for High-Performance Computing, Networking, Storage and Analysis* (2014).

S. Rajbhandari, A. Nikam, P.-W. Lai, K. Stock, S. Krishnamoorthy, and P. Sadayappan. 2014. CAST: Contraction algorithm for symmetric tensors. *Proceedings of the 2014 43rd International Conference on Parallel Processing,* (2014). DOI:10.1109/ICPP.2014.35.

B. Sanders, R. Bartlett, E. Deumens, V. Lotrich, and Mark Ponton. 2010. A block-oriented language and runtime system for tensor algebra with very large arrays. *Proceedings of the 2010 ACM/IEEE International Conference for High-Performance Computing, Networking, Storage, and Analysis*, IEEE Computer Society, Washington DC, ISBN 978-1-4244-7558-9.

Angelo Signoracci, T. Duguet, G. Hagen, and G. R. Jansen. 2015. Ab initio Bogoliubov coupled-cluster theory for open-shell nuclei. *Physical Review C* 91, (2015), 064320. DOI: 10.1103/PhysRevC.91.064320.

Jaewoong Sim, Aniruddha Dasgupta, Hyesoon Kim, and Richard Vuduc. 2012. A performance analysis framework for identifying potential benefits in GPGPU applications. In *Proceedings of the 17th ACM SIGPLAN symposium on Principles and Practice of Parallel Programming* (PPoPP '12). ACM, New York, NY, USA, 11-22. DOI=http://dx.doi.org/10.1145/2145816.2145819

E. Solomonik, D. Matthews, J. R. Hammond, J. Demmel. 2013. Cyclops tensor framework: Reducing communication and eliminating load imbalance in massively parallel contractions. *Technical Report No.* UCB/EECS-2013-11, (2013): http://www.eecs.berkeley.edu/Pubs/TechRpts/2013/EECS-2013-11.html

Paul Springer, Jeff R. Hammond, and Paolo Bientinesi. 2016. TTC: A high-performance Compiler for Tensor Transpositions.

Paul Springer, Aravind Sankaran, and Paolo Bientinesi. 2016. TTC: A Tensor Transposition Compiler for Multiple Architectures.

Paul Springer and Paolo Bientinesi. 2016. Design of a high-performance GEMM-like tensor-tensor multiplication. *arXiv*:1607:00145 (2016).

David S. Wang, Charles D. Hill, and Lloyd C. L. Hollenberg. 2015. Simulations of Shor's algorithm using matrix product states. *arXiv*:1501.07644 (2015).

Deborah A. Weighhill and Daniel A. Jacobson. 2015. 3-way networks: Application of hypergraphs for modeling increased complexity in comparative genomics. *PLoS Computational Biology* 11, (2015), e1004079.